COOL DWARFS IN WIDE MULTIPLE SYSTEMS

PAPER 4: A COMMON-PROPER-MOTION PAIR OF TWO IDENTICAL MID-M DWARFS SEPARATED BY ABOUT 10000 AU

By F. M. Rica
*Agrupación Astronómica de Mérida, Extremadura, Spain*
*Liga Iberoamericana de Astronomía, Santa Fe, Argentina*

and José A. Caballero
*Centro de Astrobiología (CSIC-INTA), Madrid, Spain*

LSPM J0651+1843 and LSPM J0651+1845 are two high-proper-motion stars recently claimed to form a binary system, FMR 83, by Rica[1]. Here we characterize the system in detail, using astrometric and photometric data, and find that the pair consists of two M4±1 dwarfs separated by $9500^{+6200}_{-3800}$ AU. With these results, FMR 83 becomes one of the very few 'ultrafragile' systems (*i.e.*, systems with very low total masses and very wide physical separations), many of which have been identified in this series of papers.

*Two identical mid-M dwarfs separated by about 10000 AU*

Rica[1] presented 173 new wide binaries found in the Lépine & Shara[2] catalogue of high-proper-motion northern stars with $\mu > 150$ mas yr$^{-1}$. Of them, four were pairs of M dwarfs with estimated physical separations of over 5000 AU and very low total masses ($\mathcal{M}_A + \mathcal{M}_B < 0.4$ M$_{sol}$). Here, we study in detail the most promising binary in Rica[1] for the investigation of cool dwarfs in the least bound systems[3].

The binary itself is FMR 83, which is formed by LSPM J0651+1843 (hereater FMR 83 A) and LSPM J0651+1845 (FMR 83 B). *VizieR* and *Simbad* only provide their coordinates, magnitudes in the photographic blue $B_J$, red $R_F$, infrared $I_N$ and 2MASS *JHKs* passbands and proper motions (identical, of about 316 mas yr$^{-1}$) compiled by Lépine & Shara[2]. Additionally, Rica[1] provided rough estimations of spectral type, absolute magnitude in *V* band, distance, tangential velocity and mass for both components. He identified the pair using data in Lépine & Shara[2], verified its common-proper-motion with an RGB composition of Digital Sky Survey images built with the *Aladin* interactive sky atlas[4] and, finally, confirmed the common-proper-motion using two astrometric epochs separated by 3.8 yr. He estimated a projected physical separation of 9800 AU and a total mass of 0.36 M$_{sol}$. Although the derived data are correct in principle, here we revisit the system for several reasons: (*i*) There are not many known binaries with such wide separations and low total masses. Indeed, it may be one of the least bound known binaries (Paper 1 of this series[3] and references therein). (*ii*) *V* magnitudes from Lépine & Shara[2], in the absence of *Tycho-2* $V_T$ data, were estimated from photographic $B_J$ and $R_F$ magnitudes, which have in general large uncertainties (of over 0.4 mag in some cases). (*iii*) Spectral-type estimations from *V–J* colours as in Lépine & Gaidos[5] have turned out to provide types that are inaccurate by one subtype in average and up to four or five subtypes in extreme cases when compared with real low-resolution spectra (Alonso-Floriano *et al.*, in prep. — the authors carry out a detailed characterization of the M-dwarf input catalogue for CARMENES[6]). (*iv*) Rica[1] tabulated different distances for both FMR 83 A and B (78 and 61 pc, respectively). In spite of the wide separation, if physically bound, they should be located at the same distance. (*v*) Photometrically-estimated spectral types in Rica[1] were M4V and M4.5V. However, magnitudes of both FMR 83 A and B are different by less than 0.1 mag from 0.4 to 11 μm (see below), which indicates identical spectral types.

We carried out an analysis similar to that in previous issues of this series (*e.g.*, Paper 3[7]). First, we used the imaging instrument CAMELOT (2k x 2k, 0.305 arcsec/px) at the 0.8 m IAC-80 telescope on Teide Observatory, Tenerife, to obtain optical images of FMR 83 AB on 28 Feb 2012 in the Johnson *BVRI* bands and on 29 Feb 2012 in the SDSS *g'r'i'z'* bands. We used the AAVSO Photometric All-Sky Survey (APASS, http://www.aavso.org/apass) to calibrate the *B*, *V*, *g'*, *r'* and *i'* images (the eighth data release of the Sloan Digital Sky Survey[8] did not cover the area) and the Third US Naval Observatory CCD Astrograph Catalogue[9] (UCAC3) to calibrate the *R* image. The *I* image was calibrated with photometric standard stars observed at different air masses. We were not able to find a zero-magnitude for calibrating our *z'* image, but we measured $\Delta z'$ = 0.05±0.01 mag on it (A brighter than B).

Our photometry, provided in Table I, is consistent with and dramatically enhances published photographic plate- (USNO-A2 $B_J R_F$, USNO-B1 $I_N$) and CCD-based (CMC14 *r'*, UCAC3 *R*) magnitudes. (Only the Carlsberg Meridian Catalogue 14[10] tabulates accurate magnitudes for both FMR 83 A and B). We complemented our data with infrared photometry from the Two-Micron All Sky Survey[11] and *WISE* Preliminary Data Release[12] (the *WISE W1*, *W2*, *W3* and *W4* bands correspond to 3.4, 4.2, 11.6 and 22.1 μm, respectively; tabulated magnitudes are the average of two independent measurements).

Next, we quantified the constancy of the angular separation during eleven astrometric epochs spread over a time interval of over 60 years. Most of the astrometric epochs corresponded to digitized images of old photometric plates (of the Digital Sky Survey). We did not use the STScI QuickV digitization of the 1983 Jan 13 plate because of its apparent astrometric offset with respect to the rest of plates. Our proper-motion determination, shown in Table I, matches that of Lépine & Shara[2] within uncertainties. In Table II, we provide observation date, angular separation, position angle and origin of every astrometric measurement. FMR 83 B felt between two relatively blue background sources at the end of the 1990s and early 2000s, but this fact did not affect the relative astrometry. During the observed 60.3 yr, we measured average constant angular separation of 111.85±0.10 arcsec (1.86 arcmin) and position angle of 330.69±0.09 deg, thus confirming the memberhip in a common-proper-motion pair of both FMR 83 A and B.

By comparing the *r'-i'* and *i'-J* colours of the two stars with those of M dwarfs tabulated by West *et al.*[13], we estimated a spectral type M4±1 V for both FMR 83 A and B. From the apparent *J* magnitude and the absolute *J* magnitude-spectral type relation by Caballero et al.[14], we conservatively estimated an heliocentric distance of $d = 80^{+60}_{-30}$ pc. At this distance, the angular separation translates into a projected physical separation of $s = 9500^{+6200}_{-3800}$ AU ($10^{+6}_{-4}$ kAU). These values are in agreement, within error bars, with those tabulated by Rica[1]. Dwarfs of spectral type M4±1 have masses of $\mathcal{M} = 0.25\pm0.07$ $M_{sol}$ (A. Reiners, priv. comm.), which makes the system to have a total mass of $\mathcal{M}_{total} = 0.50\pm0.10$ $M_{sol}$.

Finding common-proper-motion stars separated by tens of thousands of astronomical units is not extraordinary: for example, Proxima Centauri, the closest star to our Sun, is located at a physical, non-projected, separation of 12000±600 AU to α Cen A. However, the low total mass of the FMR 83 system, about four times lower than the triple α Cen system, is remarkable. This makes the FMR 83 gravitational potential energy, $U^*_g$, computed as in Caballero[15], to be as low as –(10-12) $10^{33}$ J. This value is comparable to those of the exclusive trio of very wide ($s > 1000$ AU), very low-mass ($\mathcal{M}_{total} \ll 1$ $M_{sol}$), equal-mass ($q = \mathcal{M}_B / \mathcal{M}_A \sim 1$), 'ultrafragile' binaries: Koenigstuhl 1[16], 2M0126–50[17] and 2M1258+40[18]. FMR 93 and Koenigstuhl 4, another wide low-mass binary that was discussed in Paper 1[3], have slightly larger binding energies and higher total masses than the trio. Thus, they fill the gap in the binding-energy–total-mass diagram[15] between the fragile triad and the rest of resolved binaries of ultracool dwarfs with projected physical separations to which astronomers were used in the past.

*Acknowledgements*

Financial support was provided by the Spanish MICINN under grants RYC2009-04666 and AYA2011-30147-C03-03. This article is based on observations with the IAC-80 telescope operated by the Instituto de Astrofísica de Canarias in the Spanish Observatorio del Teide. We thank R. Barrena and J. García-Rojas for carrying out observations for us in service time.

Table I
*Basic data of FMR 83 A and B*

| Datum | A | B | Origin |
|---|---|---|---|
| Name | LSPM J0651+1843 | LSPM J0651+1845 | 1 |
| $\alpha$ (J2000) | 06 51 04.51 | 06 51 00.65 | 10 |
| $\delta$ (J2000) | +18 43 42.2 | +18 45 19.7 | 10 |
| $\mu_\alpha \cos\delta$ [mas a$^{-1}$] | +203±2 (+205) | +201±3 (+205) | This work (1) |
| $\mu_\delta$ [mas a$^{-1}$] | −241±2 (−241) | −242±3 (−241) | This work (1) |
| B [mag] | 18.96±0.03 | 19.06±0.03 | This work |
| g' [mag] | 18.06±0.02 | 18.05±0.02 | This work |
| V [mag] | 17.50±0.05 | 17.48±0.05 | This work |
| R [mag] | 16.86±0.08 | 16.76±0.08 | This work |
| r' [mag] | 16.79±0.05 | 16.84±0.05 | This work |
| i' [mag] | 15.16±0.02 | 15.16±0.02 | This work |
| I [mag] | 14.49±0.10 | 14.42±0.10 | This work |
| J [mag] | 12.983±0.023 | 12.981±0.021 | 11 |
| H [mag] | 12.527±0.023 | 12.484±0.025 | 11 |
| $K_s$ [mag] | 12.216±0.018 | 12.210±0.020 | 11 |
| W1 [mag] | 12.05±0.03 | 12.01±0.02 | 12 |
| W2 [mag] | 11.82±0.03 | 11.81±0.03 | 12 |
| W3 [mag] | 11.45±0.20 | 11.31±0.19 | 12 |
| W4 [mag] | > 9.0 | > 8.4 | 12 |
| Phot. Sp. Type | M4: V | M4: V | This work |

Table II
*Astrometric observations of the FMR 83 AB system*

| Epoch | $\rho$ [arcsec] | $\theta$ [deg] | Origin |
|---|---|---|---|
| 1951 Nov 08 | 111.94 | 330.65 | POSSI Blue |
| 1951 Nov 08 | 111.91 | 330.54 | POSSI Red |
| 1989 Nov 03 | 111.72 | 330.66 | DSS 1989.843 |
| 1990 Dec 17 | 111.79 | 330.85 | POSSII Red |
| 1993 Nov 28 | 111.92 | 330.75 | DSS 1993.857 |
| 1993 Dec 05 | 111.89 | 330.69 | POSSII Infrared |
| 1993 Dec 10 | 112.01 | 330.77 | DSS 1993.942 |
| 1995 Oct 27 | 111.92 | 330.80 | POSSII Blue |
| 1997 Nov 28 | 111.71 | 330.63 | 2MASS[11] |
| 2001 Aug 31 | 111.83 | 330.65 | CMC14[10] |
| 2012 Feb 29 | 111.72 | 330.65 | This work [CAMELOT] |
| *Average* | 111.85±0.10 | 330.69±0.09 | This work |

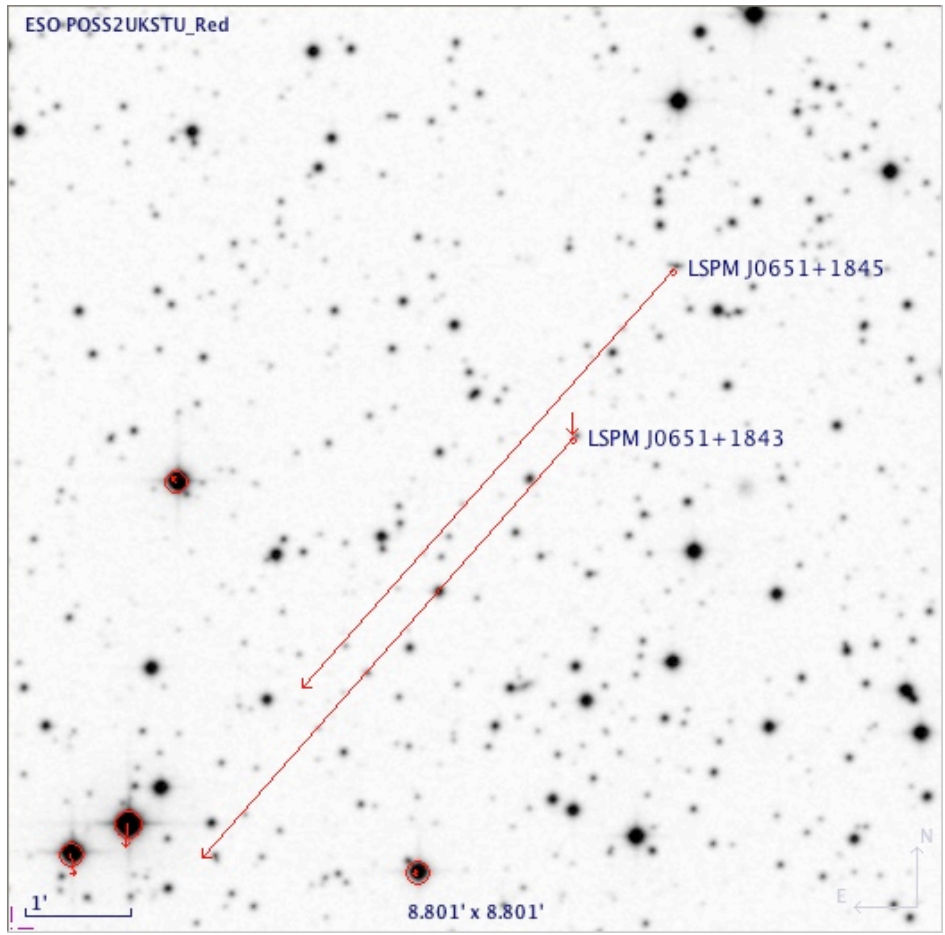

FIG. 1

Colour-inverted $R_F$-band image of *The Digitized Sky Survey* constructed with Aladin showing FMR 83 A (LSPM J0651+1843, southeast) and B (LSPM J0651+1845, northwest), which are labelled. The long arrows indicate the proper motions of stars as tabulated by *Simbad*. North is up, east is to the left. Approximate field of view is 8.8 × 8.8 arcmin².